\documentclass[manuscript]{aastex}
\usepackage{emulateapj5}
\usepackage{apjfonts}
\font\mitt=cmmi9 scaled\magstep 1

\slugcomment{The Astrophysical Journal, draft version \today}

\def\gtsima{$\; \buildrel > \over \sim \;$}
\def\ltsima{$\; \buildrel < \over \sim \;$}
\def\gsim{\lower.7ex\hbox{\gtsima}}
\def\lsim{\lower.7ex\hbox{\ltsima}}
\def\simgt{\lower.7ex\hbox{\gtsima}}
\def\simlt{\lower.7ex\hbox{\ltsima}}
\def\la{\lsim}

\def\HI{\ifmmode \hbox{\scriptsize H\kern0.5pt{\footnotesize\sc i}}\else H\kern1pt{\small I}\fi}
\def\Halpha{H$\alpha$}
\def\mlstar{\ifmmode\Upsilon_{\!\!*}\else$\Upsilon_{\!\!*}$\fi}
\def\mlstarR{\ifmmode\Upsilon_{\!\!*}^R\else$\Upsilon_{\!\!*}^R$\fi}
\def\kms{\ifmmode\mathrm{~km~}\mathrm{s}^{-1}\else km s$^{-1}$\fi}
\def\cm2{cm$^{-2}$}
\def\pc2{pc$^{-2}$}
\def\pc3{pc$^{-3}$}
\def\varv{\hbox{\mitt v}}

\def\lab{\rlap{\raise2pt\hbox{$<$}}\lower2.5pt\hbox{$\sim$}}
\def\gab{\rlap{\raise2pt\hbox{$>$}}\lower2.5pt\hbox{$\sim$}}

\newcommand{\tskip}{\omit\tablevspace{1pt}}

\shorttitle{Core Kinematics of the LSB Galaxy DDO 39}
\shortauthors{Swaters, Verheijen, Bershady, Andersen}

\begin{document}

\title{The Kinematics in the Core of the Low Surface Brightness
Galaxy DDO 39}

\author{R. A. Swaters\altaffilmark{1,2},
M. A. W. Verheijen\altaffilmark{3,4},
M. A. Bershady\altaffilmark{3},
D. R. Andersen\altaffilmark{5}}

\altaffiltext{1}{Department of Physics and Astronomy, Johns Hopkins
University, 3400 N. Charles Str., Baltimore, MD 21218, and Space
Telescope Science Institute, 3700 San Martin Dr., Baltimore, MD
21218. $^2$ Visiting Astronomer, Kitt Peak National Observatory,
National Optical Astronomy Observatory, which is operated by the
Association of Universities for Research in Astronomy, Inc. (AURA)
under cooperative agreement with the National Science Foundation. The
WIYN Observatory is a joint facility of the University of
Wisconsin-Madison, Indiana University, Yale University, and the
National Optical Astronomy Observatory. $^3$ Astronomy Department,
University Wisconsin - Madison, 475 N. Charter St., Madison, WI
53706. $^4$ Astrophysikalisches Institut Potsdam, An der Sternwarte
16, 14482 Potsdam, Germany. $^5$ Max Planck Institut f\"ur Astronomie,
K\"onigstuhl 17, 69117 Heidelberg, Germany}

\begin{abstract}
  We present a high resolution, SparsePak two-dimensional velocity
  field for the center of the low surface brightness (LSB) galaxy
  DDO~39.  These data are a significant improvement on previous \HI\
  or \Halpha\ long slit data, yet the inner rotation curve is still
  uncertain due to significant noncircular and random motions. These
  intrinsic uncertainties, probably present in other LSB galaxies too,
  result in a wide range of inner slopes being consistent with the
  data, including those expected in cold dark matter (CDM)
  simulations.  The halo concentration parameter provides a more useful test of
  cosmological models than the inner slope as it is more tightly
  constrained by observations.  DDO~39's concentration parameter
  is consistent with, but on the low end of the distribution predicted
  by CDM.
\end{abstract}

\keywords{ galaxies: dwarfs --- galaxies: halos --- galaxies:
  kinematics and dynamics }

\section{Introduction} 
\label{secintro}

Because low surface brightness (LSB) galaxies are most likely
dominated by dark matter at all radii (e.g., de Blok \& McGaugh 1997;
Verheijen 1997; Swaters, Madore, \& Trewhella 2000), they are ideal
for studying the power law slopes of the central dark matter density
distributions $\rho(r)\propto\rho^{-\alpha}$.  Cosmological
simulations indicate $\alpha$ depends on the nature of the dark matter
(e.g., Navarro, Frenk, \& White 1997, hereafter NFW; Fukushige \&
Makino 1997; Moore et al.\ 1999; Col\'{\i}n, Avila-Reese, \&
Valenzuela 2000; Dav\'e et al.\ 2001; Knebe et al.\ 2001). Thus, a
measurement of $\alpha$ may provide constraints on the nature of dark
matter in galaxies and theories of galaxy formation.

Unfortunately, it is difficult to measure the inner slope $\alpha$
observationally. \HI\ observations have relatively low angular
resolution and are affected by beam smearing (Swaters 1999, hereafter
S99; Swaters et al.\ 2000; van den Bosch et al.\ 2000; McGaugh, Rubin,
\& de Blok 2001); hence \HI\ observations generally are less
suited to measure $\alpha$ accurately. \Halpha\ long-slit observations
provide higher angular resolution, but studies based on such
observations find conflicting results.  Some find LSB galaxies are
consistent with steep inner slopes (Swaters 2001; Swaters et al. 2003,
hereafter SMvdBB), yet others find them inconsistent with steep
slopes (de Blok et al. 2001a,b; de Blok \& Bosma 2002, hereafter dBB;
Marchesini et al.\ 2002). This inconsistency remains even when
identical datasets are used (SMvdBB).  The apparent discrepancy
can be explained, in part, by systematic effects of slit width,
seeing, slit offsets and galaxy inclination, which all lead to an
underestimate of $\alpha$ (SMvdBB).

\begin{figure*}[th]
\begin{center}
\epsfxsize=1.0\hsize \epsfbox{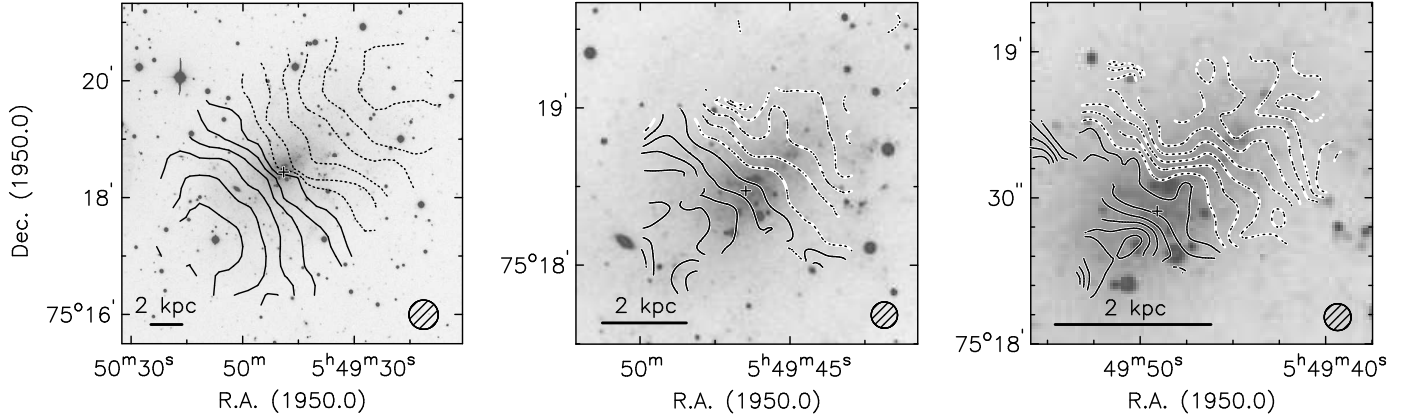}
\vskip0pt
\figcaption{\HI\ velocity field from Swaters et al. (2002) (left
panel), SparsePak $10''$ velocity field (middle panel), and SparsePak
$5.5''$ velocity field (right panel) superposed on the optical image
from Swaters \& Balcells (2002). Steps between isovelocity contours
are 10 \kms\ in the left two panels, and 5 \kms\ in the right
panel. The dotted lines represent the approaching side, the first full
line indicates the systemic velocity. The beam size is given in the
lower right of each panel, the cross indicates the galaxy center. Most
bright objects are foreground stars.\label{figvfs}} 
\vspace*{-0.5cm}
\end{center}
\end{figure*}

To avoid systematic effects of long-slit observations, and to map
possible non-circular motions, it is necessary to obtain high spatial
resolution, two-dimensional velocity fields (see also Beauvais \&
Bothun 1999). Blais-Ouellette et al.\ (2001) presented such
observations for the dwarf LSB galaxies NGC~3109 and
IC~2574. Unfortunately, both galaxies are less suited to address the
core issue; NGC~3109 is close to edge-on, and IC~2574 has a poorly
sampled velocity field and a perturbed interstellar medium (Walter \&
Brinks 1999).

In this Letter, we present a high-resolution, two-dimensional velocity
field of the LSB galaxy DDO~39. This galaxy was selected because of
its suitable inclination of $49^\circ$, low surface brightness
($\mu_B=24.4$ mag arcsec$^{-2}$, Swaters \& Balcells 2002), disk
dominated radial surface brightness profile with a disk scale length
of $h=3.5$~kpc at an adopted distance of 12.8 Mpc (S99), regular
optical and \HI\ morphology, and well-behaved \HI\ kinematics (see
Fig.~\ref{figvfs}). These properties make DDO~39 an ideal galaxy to
study the kinematics in the central regions of an LSB galaxy.

\section{Observations and data reduction}
\label{secobs}

DDO~39 was observed with the SparsePak integral field unit on the 3.5m
WIYN telescope on January 11, 12 and 13, 2002.  SparsePak is a
fiber-optic array containing 82 fibers, each $4.7''$ in diameter and
separated by $5.6''$, arranged in a sparsely packed grid, $72''$ on
each side, with a small, nearly-integral core. Seven sky fibers are
spaced between $60''$ and $90''$ away from the central fiber.  The
grid can be fully sampled in three pointings.  For a detailed
description, see Bershady et al.\ (in preparation). The fibers are fed
to the WIYN Bench Spectrograph. We used the 860 l/mm grating in second
order at 6600 \AA, yielding a FWHM velocity resolution of 66 \kms.  In
the first two nights we filled the grid with three $3\times 20$~min
integrations; in the third night we integrated $3\times 1$~hour on the
central pointing.

The IRAF dohydra package was used for data reduction and extraction of
the spectra. Sky subtraction was done by creating a two-dimensional
image of extracted spectra, filtering out the H$\alpha$ lines, and
subtracting a low order baseline (Bershady et al, in preparation). To
determine the actual telescope pointing and offsets, we calculated the
expected relative continuum flux for all fibers from an $R$-band image
of DDO~39, and compared this to our observations. By minimizing the
$\chi^2$ difference, the pointing offsets could be determined with an
accuracy of $\sim 0.5''$ (Swaters et al., in preparation).

We constructed a sparse velocity field by fitting Gaussians to spectra
with \Halpha\ emission stronger than three times the noise level and
placing the derived velocities in a map at the corresponding
positions. To construct a contiguous velocity field for graphical
presentation, the points in the sparse velocity field were
interpolated after weighting with a Gaussian beam of $5.5''$ and
$10''$. These velocity fields, shown in Fig.~\ref{figvfs}, highlight
the kinematics in the central regions and over the entire SparsePak
field of view. Because of the interpolation, the isovelocity contours
may be uncertain, especially at the edges.

\section{Results}
\label{secderrcs}

\subsection{The rotation curve}

To determine the galaxy orientation parameters, we fitted a tilted
ring model to the sparse velocity field.  Because of the approximately
linear rise of the rotation curve (RC), it was not possible to
determine the center or the inclination kinematically from the
SparsePak data.  Instead, we used the optical center at $5^h 56^d
37.5^s$ $75^\circ 18' 56''$ (J2000). Although the inner rotation curve
shape depends somewhat on the choice of center, the main results of
this Letter do not. For the inclination we used $49^\circ$, derived
from the \HI\ observations.  The position angle and systemic velocity
were found to be consistent with the $133^\circ$ and 818 \kms\ of the
\HI\ observations, and we used the latter.

With these orientation parameters, all points in the sparse velocity
field were corrected for projection effects. To determine the rotation
velocity, points were averaged in concentric annuli with a width of
$5''$. Each point was assigned a weight of
$(\cos\phi/\sigma_\mathrm{rad})^2$, where $\phi$ is the angle in the
plane of the galaxy with respect to the major axis, and
$\sigma_\mathrm{rad}$ is the error on the radial velocity, with an
imposed minimum of 4 \kms. The error on the rotation velocity was
taken to be the quadratic sum of the formal error and one fourth of
the difference between the approaching and receding sides, with an
imposed minimum error of 2 \kms.  The RC derived in this
way was combined with the \HI\ RC from S99 to create the
hybrid RC shown in Fig.~\ref{figrc}a.

Fig.~\ref{figrc}a shows good agreement between our RC and the RC
derived by S99 from \HI\ data. Fig.~\ref{figrc}b compares dBB's
long-slit data and our data within $2.5''$ of the major axis. There is
good general agreement, although the velocities derived by dBB tend to
be somewhat closer to the systemic velocity. Fig.~\ref{figrc}c
compares our RC to the one derived by dBB. Although consistent within
the errors, their RC is substantially lower where independent optical
data are used (dBB also used S99's data beyond $\sim90''$). The reason
for this is unclear, but the fact that we used two-dimensional data
and dBB spline-interpolated their binned data to derive their RC are
both likely to play a role.

\subsection{Noncircular motions}

Above we have assumed that the gas moves on circular orbits. However,
the \Halpha\ velocity fields in Fig.~\ref{figvfs} reveal significant
noncircular motions.  Because the velocity fields might be affected by
interpolation, we have also plotted the data directly. In
Fig.~\ref{figrings} we plot the observed radial velocities with
respect to systemic velocity in different radial intervals, normalized
by the rotation velocity. At large radii the radial velocities follow
the curve expected for circular rotation closely. On the other hand,
in the innermost regions there are clear deviations from simple
circular motions.

Fig.~\ref{figrings} also shows that there is significant scatter
around the best fit rotation velocity. The scatter (before
normalization) is around 9 \kms, independent of radius. The scatter is
in part due to observational uncertainties. Measurement errors depend
on the signal-to-noise ratio and range from 2 to around 8 \kms\ for
the weakest lines, and average about 4 \kms. Additional scatter may
arise if the intensity weighted position of the \Halpha\ over the
fiber ``beam'' does not coincide with the center of that fiber.  The
maximum contribution of this effect, modeled by calculating the
extreme velocities in each fiber and comparing those to the expected
velocities, is found to be lower than 3 \kms\ at all radii.
Subtracting the dominant sources of observational uncertainties from
the observed dispersion in the rings, we find the intrinsic
fiber-to-fiber line-of-sight dispersion is around 8 \kms.

Because the dispersion $\sigma$ is comparable to the rotation velocity
$\varv_\phi$ in the central regions, it may be necessary to correct
the RC to obtain the circular velocity.  If one assumes the dispersion
represents pressure support due to random motions, then, assuming
axisymmetry and absence of radial motions, the correction can be
calculated and is found to be negligible (see e.g., S99).
Alternatively, if one assumes the central regions are in a steady
state, one can calculate the circular velocity $\varv_c$ using the
virial theorem, $2T+\Pi=W$. Although the kinetic energy in rotation
$T$ and in random motions $\Pi$ depend on the distribution of the gas,
and the gravitational potential $W$ depends on the total mass
distribution, comparison to more detailed calculations\break

\hskip-0.3cm\resizebox{0.99\hsize}{!}{\includegraphics{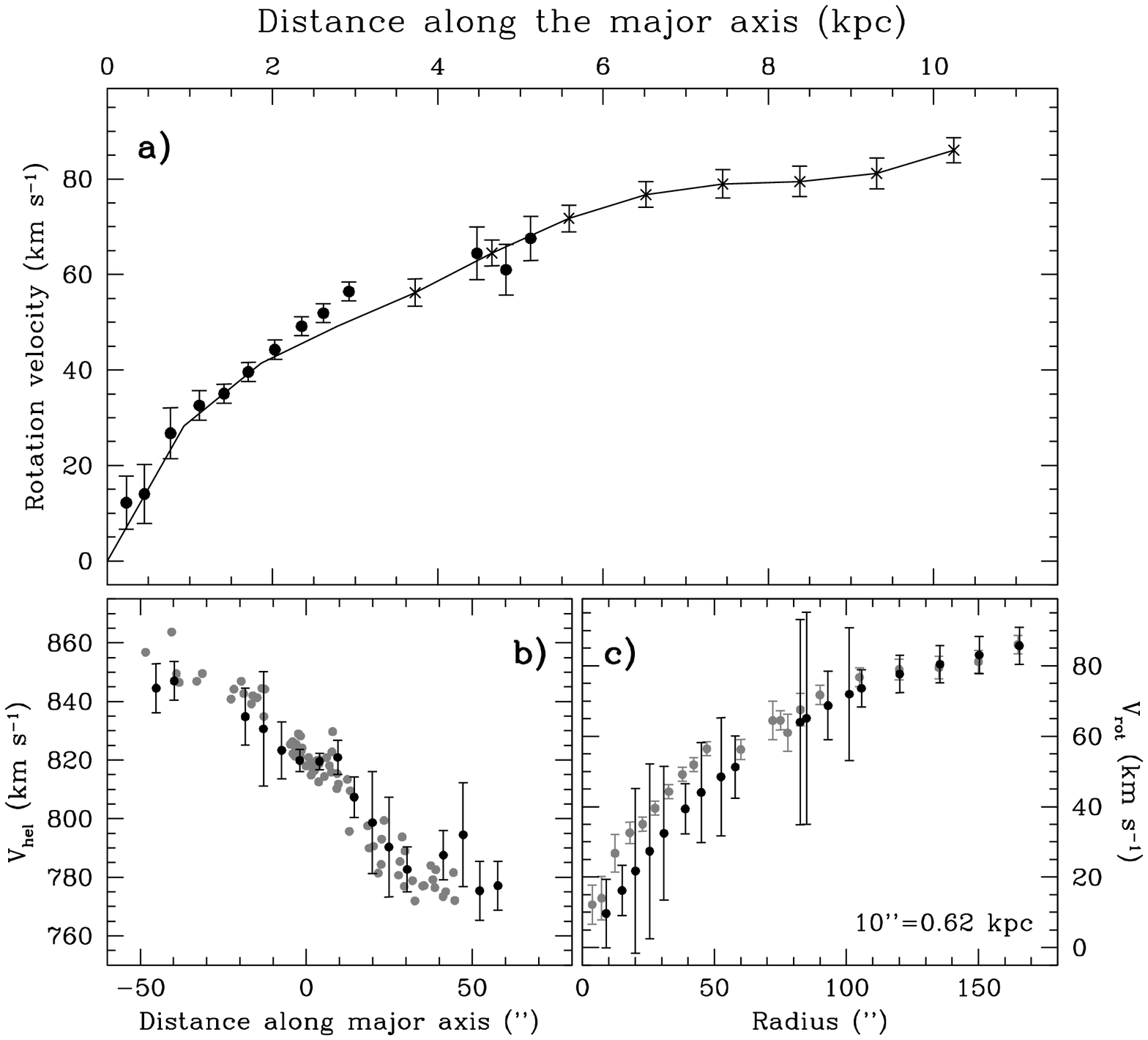}}
{\small {\sc Fig.~\ref{figrc}.}---
  {\bf a)} Hybrid rotation curve from SparsePak data ({\it dots}) and
the \HI\ data from S99 ({\it crosses}). The rotation curve from S99 is
given by the solid line. {\bf b)} Comparison of dBB's data ({\it black
dots}) to SparsePak data within $2.5''$ of the major axis ({\it grey
dots}). {\bf c)} Comparison of dBB's and our rotation curve, coding as
in panel b.
}
\bigskip
\addtocounter{figure}{1}

\noindent show the
circular velocity is approximately equal to:
\begin{equation}
\varv_c^2=\varv_\varphi^2+\beta\sigma^2.
\label{eqvirial}
\end{equation}
\noindent Assuming an isotropic dispersion ($\beta=3$) results in a
steeper RC (see Fig.~\ref{figmassmodels}).

Because the origins of the dispersion and noncircular motions are not
known, and because the assumptions for the corrections may not be
valid, these corrections are uncertain and merely give an indication
of the possible range.

\subsection{Mass models}

The contribution of the stellar disk to the RC was
calculated from the $R$-band light profile presented in Swaters \&
Balcells (2002), and the contribution of \HI\ was calculated from the
\HI\ surface density profile presented in Swaters et al. (2002),
scaled up by a factor of 1.32 to account for helium. The stellar disk
was assumed to have a vertical sech-squared distribution with a scale
height $z_0=h/6=0.58$~kpc, and the \HI\ disk was assumed to be
infinitely thin. For the dark matter halo we considered a generalized
NFW halo of the form:
\begin{equation}
\label{eqgennfw}
\rho(r) = {\rho_0 \over (r/r_s)^{\alpha} (1 + r/r_s)^{3-\alpha}},
\end{equation}
where $r_s$ is the scale radius and $\rho_0$ the central
density. These two parameters are linked to the more commonly used
parameters $c$ and $\varv_{200}$, depending on cosmology (see
Section~\ref{secdisc}).

We have fit mass models over a range of $\alpha$ and mass-to-light
ratios (M/Ls). In each fit $\alpha$ and the M/Ls were kept fixed and
the halo parameters were allowed to vary.  Confidence levels were not
calculated because the velocities and their errors are not free of systematic effects, and hence the $\chi^2$ values were
only used for relative comparison between the models.  In
Fig.~\ref{figgenfits} we plot the resulting $\chi^2_r$ values as a
function of $\alpha$ for an M/L of 1. Fits to the uncorrected and
pressure corrected RCs give virtually identical results, with best
fits in the range $0<\alpha\la 0.8$. For\break

\hskip-0.3cm\resizebox{0.99\hsize}{!}{\includegraphics{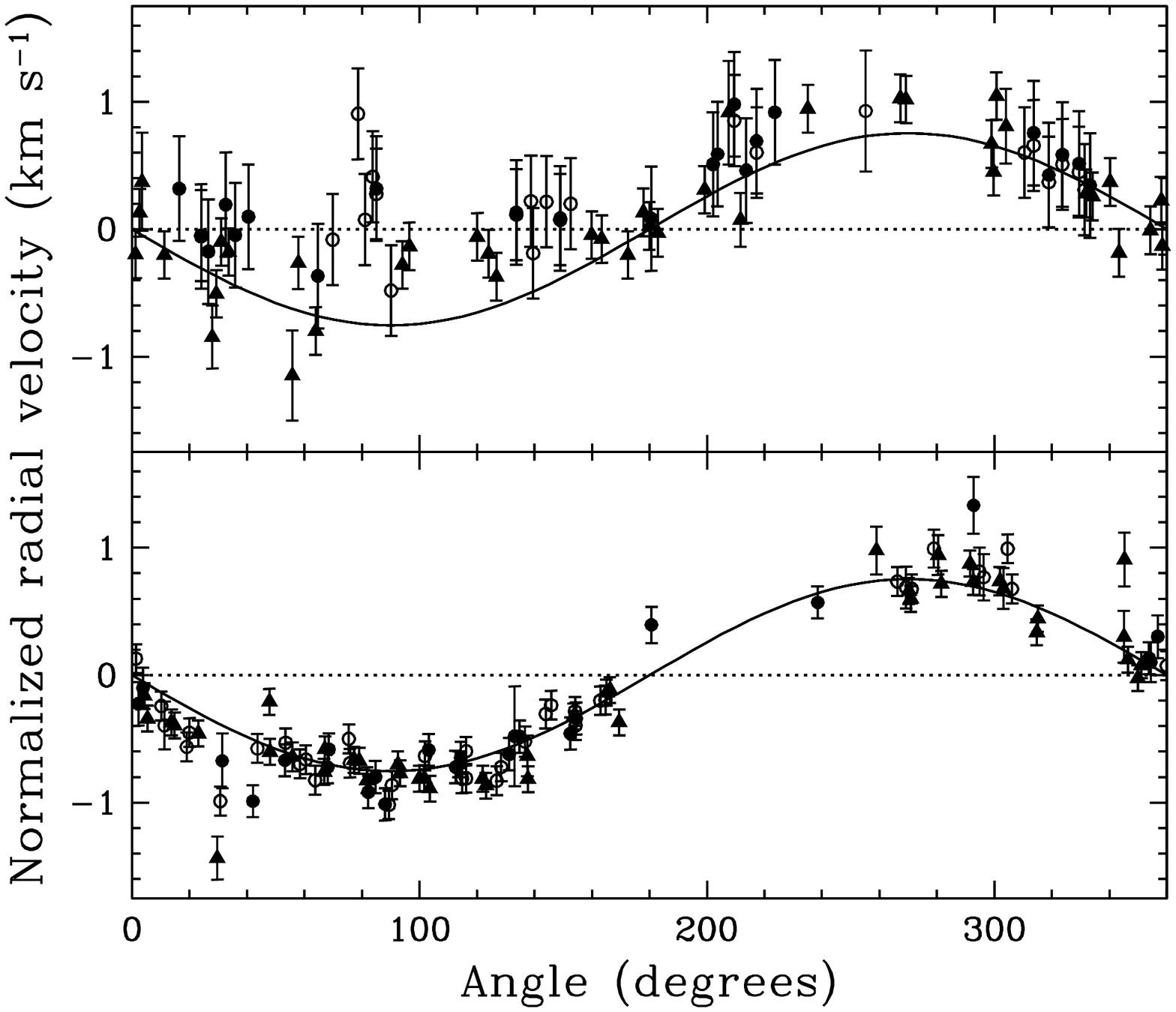}}
{\small {\sc Fig.~\ref{figrings}.}---
Radial velocity with respect to the systemic velocity, normalized by
the rotation velocity, as a function of angle in the plane with
respect to the minor axis. The dots, circles and triangles represent
the points at radii of 0.23, 0.45, and 0.77 kpc in the top panel, and
1.71, 2.03, and 2.35 kpc in the top panel, respectively. The solid
line represents the normalized curve for circular rotation.
}
\bigskip
\addtocounter{figure}{1}

\noindent the RC corrected using the
virial theorem, best fits are found for somewhat steeper inner slopes,
in the range $0.3\la\alpha\la 1.0$.  For higher values of $\alpha$ the
quality of the fits decreases rapidly.

Best fitting mass models for an M/L of 1 are shown in
Fig.~\ref{figmassmodels}, for $\alpha=1$ (the NFW profile) and
$\alpha=0$.  The difference between the $\alpha=0$ and $\alpha=1$
models are predominantly in the inner 1 kpc, where the RC is most
affected by noncircular and random motions (Fig.~\ref{figmassmodels}).
The best fitting parameters for a range of M/Ls and $\alpha=1$ are
given in Table~\ref{tabfitpars}, both for the pressure corrected RC
and the one corrected with Eq.~\ref{eqvirial} and $\beta=3$.

\section{Discussion and conclusions}
\label{secdisc}

The high-resolution velocity field of the LSB galaxy DDO~39 presented
here reveals the presence of noncircular motions in the central
regions and an intrinsic fiber-to-fiber dispersion of about 8 \kms.
The origin of either component is unclear; DDO~39 does not have a
strong bar or spiral arms, nor does it appear to be
interacting. Likely contributors are: star formation activity,
turbulence, and small scale structure in the disk or dark halo.
Irrespective of their origin, the noncircular motions and the large
$\sigma/\varv_\phi$ in the galaxy center make an accurate measurement
of the circular velocity curve difficult.

As a result, inner slopes in the range $0<\alpha\la 1$ are consistent
with the data, and DDO~39 is compatible with a wide range of dark
matter properties. Specifically, it appears that the data presented
here are consistent with the inner slopes expected in the currently
popular $\Lambda$CDM paradigm. Although earlier simulations predicted
inner slopes ranging from $\alpha=1.5$ (e.g., Fukushige \& Makino
1997, Moore et al.\ 1999) to $\alpha=1$ (e.g., NFW), a recent study by
Power et al.\ (2003) suggests that the differences in $\alpha$ are
mainly due to resolution issues, and places an upper limit to $\alpha$
of 1.2. Taylor \& Navarro (2001), based on analytical arguments, find
$\alpha$ may be as low as 0.75 at very small radii.\break

\hskip-0.3cm\resizebox{0.99\hsize}{!}{\includegraphics{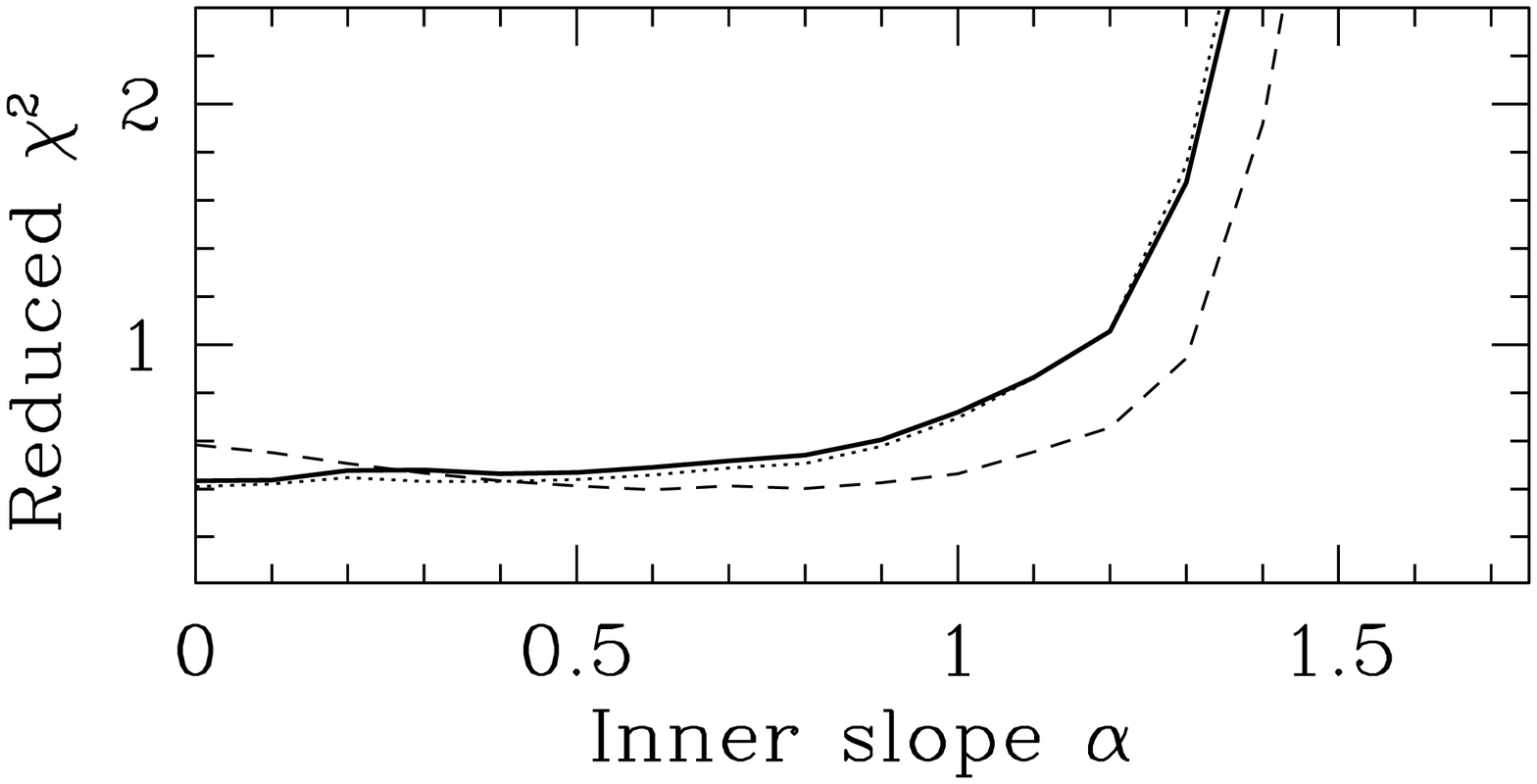}}
{\small {\sc Fig.~\ref{figgenfits}.}---
$\chi^2_r$ versus $\alpha$, as determined from fits with M/L fixed to
1, and the halo parameters left free. The full line gives the results
for fits to the uncorrected rotation curve, the dotted line for
pressure correction, and the dashed line for the virial correction of
Eq.~\ref{eqvirial}.
}
\addtocounter{figure}{1}

\begin{center}
{\sc Table \ref{tabfitpars}\\
\smallskip\hbox to\hsize{\hfil{Best fit parameters for $\alpha=1$}\hfil}}
\small
\setlength{\tabcolsep}{6pt}
\begin{tabular}{rrcrrcr}
\tskip \tableline
\tableline \tskip

& \multicolumn{3}{c}{NFW (pressure)} & \multicolumn{3}{c}{NFW (virial)} \\
\mlstarR & $c$ & $v_{200} (\kms)$ & $\chi^2_r$ & $c$ & $v_{200} (\kms)$ & $\chi^2_r$ \\
\tableline \tskip
  0     &  6.1 &    90 &  0.77 & 7.1 &  83 & 0.51 \\
  1	&  4.4 &    93 &  0.72 & 5.6 &  83 & 0.46 \\
  2	&  2.1 &   148 &  0.75 & 4.0 &  92 & 0.44 \\
  4	&  1.0 &   150 &  0.81 & 1.4 & 143 & 0.46 \\
  1	&10$^a$&    54 &   4.0 &10$^a$& 54 &  2.6 \\
\tableline
\end{tabular}
\end{center}

\centerline{\vbox{\hsize=6.5cm \footnotesize \hskip-1cm Note --- $^a$
Fit with $c$ fixed}}

\medskip

Even though the data presented here eliminate several uncertainties
that gave poor constraints on $\alpha$ from \HI\ and \Halpha\
long-slit observations (van den Bosch et al.\ 2000, van den Bosch \&
Swaters 2001, SMvdBB, but see de Blok et al.\ 2001a,b), the intrinsic
noncircular and random motions still result in poor constraints on
$\alpha$.  Given that deviations from circular motions seem common in
LSB galaxies (e.g., Walter \& Brinks 1999, de Blok \& Walter 2000),
and are present even in a very LSB galaxy such as DDO~39, it seems
that an accurate measurement of $\alpha$ and, from that, a tight limit
on the nature of dark matter may be difficult to achieve
observationally from a component that is susceptible to kinematic
perturbations.

Fortunately, the halo concentration parameter $c$ is better
constrained observationally as it is derived from a fit to the RC as a
whole rather than mostly from the central parts, although it may still
be affected by systematic effects (SMvdBB).  For the case of
$\alpha=1$, the inferred halo parameters $c$ and $\varv_{200}$ can be\break

\hskip-0.3cm\resizebox{0.99\hsize}{!}{\includegraphics{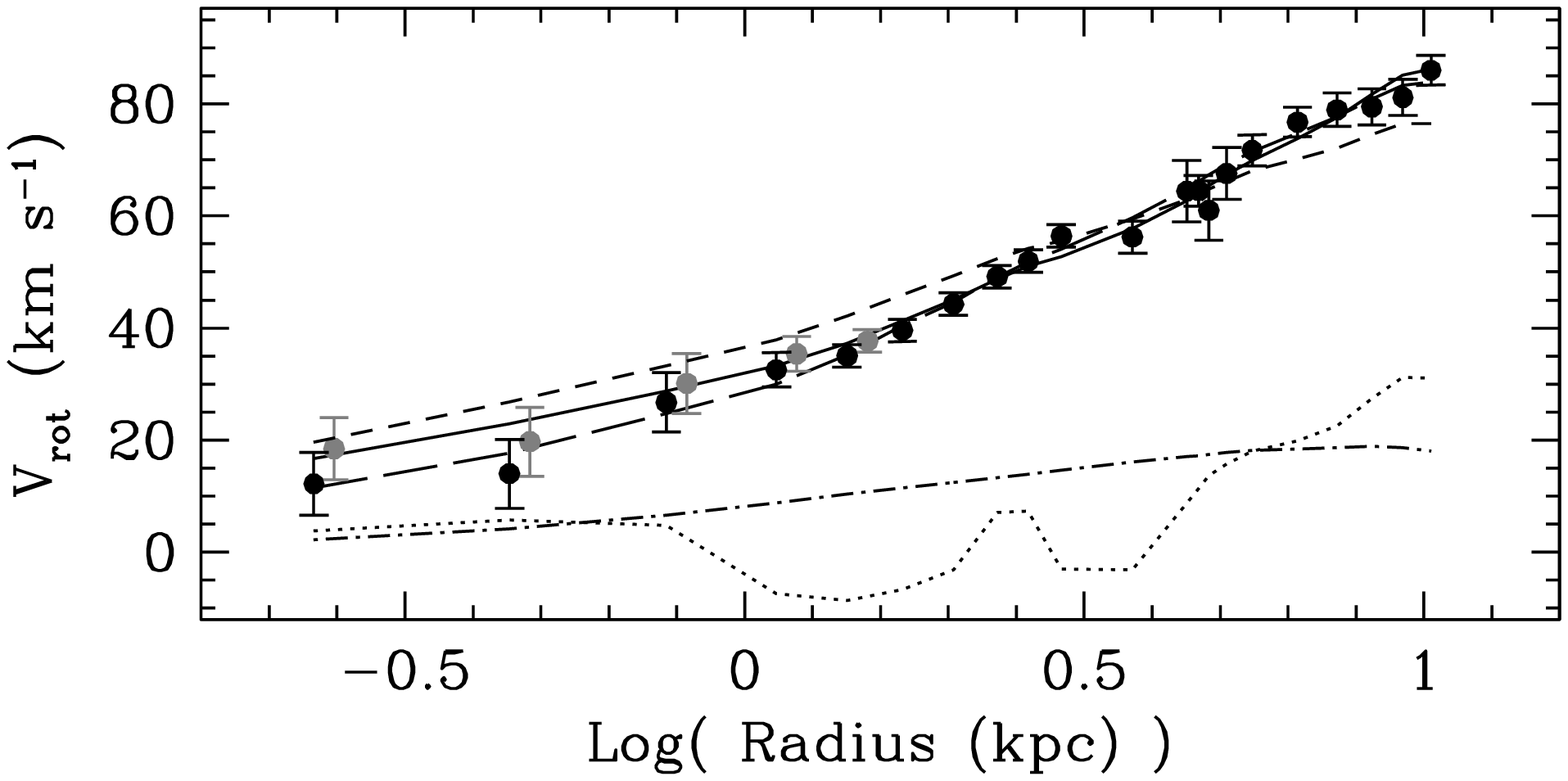}}
{\small {\sc Fig.~\ref{figmassmodels}.}---
Mass models for the best fit NFW halo (solid line), for an NFW halo
with $c=10$ (short dashed line), and for a halo with $\alpha=0$ (long
dashed line). The dotted line represents the contribution of the \HI,
the dot dashed line the contribution of the stars with an M/L of
1. The contribution of the halo is similar to the best fit model, but
is not shown. The black dots give the uncorrected rotation curve, the
gray dots (only shown for the inner five points and offset by 0.03 for
clarity) represent the rotation curve corrected using
Eq.~\ref{eqvirial} assuming $\beta=3$.
}
\bigskip
\addtocounter{figure}{1}

\noindent compared to the values expected in a $\Lambda$CDM cosmology. Expected
values have a $2\sigma$ range from 5 to 25, with an average of around
10 to 15 (NFW, Bullock et al.\ 2001). DDO~39's $c\sim 5$ is at the low
end of this distribution, and for different values of $c$ the quality
of the fit changes rapidly. For example, $c=10$ is inconsistent with
the data (see Fig.~\ref{figmassmodels} and Table~\ref{tabfitpars}).

The low value for $c$ does not necessarily indicate an inconsistency
with $\Lambda$CDM. For example, a bias toward low $c$ values in LSB
galaxies could be explained if LSB galaxies preferentially form in low
density halos. In addition, the concentration parameter also depends
on the slope of the power spectrum of density fluctuations, and $c\sim
5$ is in agreement with models in which structure formation on small
scales is suppressed (Zentner \& Bullock 2002). Furthermore, $c$ would
be higher if DDO~39 were closer than the adopted 12.8 Mpc.

Accurate RCs determined from high resolution, two-dimensional velocity
fields for a sample of LSB galaxies with intermediate inclinations,
accurate centers, and regular kinematics and morphology may provide
RCs to measure $c$, or any parameter that depends on the global shape
of the RC, with sufficient accuracy to provide useful constraints on
the nature of dark matter.

\acknowledgments

This research was supported in part from NSF grant AST-9970780.

\end{document}